\documentclass{article}
\usepackage[utf8]{inputenc}
\usepackage{graphicx}
\usepackage{caption}
\usepackage{amsmath,pdftexcmds}
\usepackage[a4paper, total={6.5in, 10in}]{geometry}
\usepackage{color}
\usepackage{braket}
\usepackage{comment}
\usepackage{abstract}
\usepackage{amsfonts}
\usepackage{subfig}
\usepackage{graphicx}
\usepackage{multirow}
\usepackage{caption}
\usepackage{pdfpages}
\usepackage{amssymb}
\usepackage{float}
\usepackage{pgfplots}
\usepackage{tikz}
\usepackage{tikz-3dplot}
\usepackage{authblk}
\usepackage{bm}
\usepackage[none]{hyphenat}
\usepackage{epsfig}
\graphicspath{graphics/}
\usepackage[colorlinks,linkcolor={black},citecolor={black},urlcolor={blue},unicode,breaklinks=true]{hyperref}

\title{\textbf{Role of mass fluctuations in the diffusion of clusters of Brownian particles with activity}}

\author{Daniela Moretti}
\author{Pasquale Digregorio}
\author{Giuseppe Gonnella}
\author{Antonio Suma}
\affil{Dipartimento Interateneo di Fisica, Universit\`a degli Studi di Bari and INFN, Sezione di Bari, via Amendola 173, Bari, I-70126, Italy. E-mail: daniela.moretti@uniba.it.}

\date{}

\pgfplotsset{compat=1.18}

\begin{document}

\maketitle

\begin{abstract}
Motivated by the anomalous diffusion observed in clusters of active Brownian particles (ABPs), where the center-of-mass diffusion coefficient scales as \(D\sim N^{-1/2}\) with respect to the number \(N\) of particles in the cluster, we derive a minimal theoretical framework starting from the single-particle Langevin equations. The model consists of two coupled stochastic equations: one for the cluster center-of-mass trajectory and one for the mass evolution \(N(t)\), explicitly accounting for stochastic displacements induced by particle attachment and detachment.
We specialize and validate the framework against ABP simulations of isolated clusters in stationary conditions, where \(N(t)\) follows a Gaussian process with mean \(N_0\), variance \(\propto N_0^\beta\), and persistence time \(\propto N_0^\kappa\). Analytical solution of the coupled equations yields the long-time diffusion coefficient as the sum of two contributions: a conventional term \(\propto N_0^{-1}\) due to thermal noise plus summation of  active forces, and a fluctuation-driven term \(\propto N_0^{-\delta}\) with \(\delta=2-2/d-\beta+\kappa\), where \(d\) is the spatial dimension. We demonstrate that anomalous scaling emerges whenever the second term becomes dominant.
The model predicts \(D\sim N^{-\alpha}\) with \(\alpha=0.63\pm0.06\), in good  quantitative agreement with large-scale ABP simulations.
\end{abstract}

\section{Introduction}
The Brownian motion of a single colloidal particle is characterized by the general Einstein expression for the diffusion coefficient $D = k_BT/\gamma m$, where $k_B$ is the Boltzmann's constant, $T$ is temperature, $\gamma$ is the damping constant, $m$ the mass of the particle.

The above relation holds true for many colloidal suspensions in dilute regimes~\cite{dhont1996introduction}. 
When extending such expression to an aggregate of $N$  Brownian particles,  the center of mass diffusion coefficient remains inversely proportional to the total mass, $D \sim 1/N$, since each bead experiences its own independent friction and uncorrelated random thermal forces, see e.g. the Rouse model of linear polymers \cite{book_theory_pol_dynamics}.

However, the dependence of $  D  $ on $  N  $ can vary upon the introduction of more complex interactions. In the Zimm model for polymer dynamics, long-range hydrodynamic correlations of the solvent modify the scaling, $  D \sim N^{-0.5}  $\cite{book_theory_pol_dynamics}. At the same time, solvent conditions and the fractal dimension of the polymer can lead to a more general scaling $  D \sim N^{-\alpha}$, with $\alpha>0$  ~\cite{auge2009nmr}.
The scaling of $  D  $ in aggregates of $  N  $ particles can also depend on the fractal dimension of the clusters \cite{meakin_clusters_93,hanggi_clusters_PRE95,rajesh2002aggregate} and on aggregation or fragmentation processes that dynamically change the mass.
The possibility that $D$ could have a distribution dependent on $N$ was introduced within a superstatistical approach~\cite{hitchhiker_model,seno_superstatistics_PRX17}, where the displacement probability was analysed under the assumption of a stationary distribution for the size $R$ of individual molecules.

A gran-canonical formalism has been also adopted for taking into account the  number of particle fluctuations on polymer dynamics, which has been found particularly suited for describing self-assembly processes or transient aggregates\cite{baldovin_polymerization_frontiers19}.

Recent research developments have shown another key factor that can affect the scaling of \(D\): the presence of an active force for each bead that induces a persistent motion and drives the system out-of-equilibrium~\cite{Marchetti2013,Bechinger2016,Eldeti_2015}.
Studies of active polymers' models~\cite{Winkler_active_review}, where each bead has an active force aligned to the backbone direction, have shown a non-trivial dependence of $D$ on both the length of the chain,  the propulsion force and the degree of alignment~\cite{Bianco_active_polymers}.
In other types of systems,
where particles align their self-propulsion direction with neighbours, the scaling exponent $  \alpha  $ varies with interaction strength~\cite{beatrici_clusters_PRE17}. 
By contrast,
clusters of Hydra endoderm cells show nearly mass-independent diffusion~\cite{lutz2024anomalous}, and
the diffusion coefficient of entangled blobs of living worms has also been measured  to be independent on the size of the blob~\cite{bonn_phase_separation_prl}.

In the context of active matter, anomalous scaling of the diffusion coefficient has also been observed for the paradigmatic model of active Brownian particles (ABPs). These are repulsive disks that self-propel with a constant force along a direction that diffuse over time. In the absence of any attractive interaction, ABPs undergo motility-induced phase separation (MIPS)~\cite{cates_mips_2008,Render_2013,bialke_lowen_mips_2013,Cates_2015,Fily_2012,Gonnella_2015,Cugliandolo_2017,digregorio_mips_2018,mips_abp_dumbells_2019} and form dense clusters which are highly dynamical in nature, i.e. they constantly change shape, open gas bubbles, grow protrusions, and exchange mass with their environment through continuous break-up and recombination events~\cite{caporusso_micromacro_2020,claudio_clusters_23}, causing shifts in the clusters center-of-mass (COM). 
By tracking the COM trajectories of individual clusters, it was recently shown that, at intermediate values of the cluster mass, the diffusion coefficient scales anomalously as
\(D \sim N^{-0.5}\)~\cite{Bechinger2016,claudio_clusters_23,Cottin-Bizonne_2018}. 
Despite this result, a theoretical model that explicitly accounts for the internal mass fluctuations of clusters of active particles, while simultaneously reproducing the observed anomalous scaling of $D$, is still missing.

Here we address this gap by deriving, starting from the single-particle Langevin dynamics of  active Ornstein–Uhlenbeck particles, a general stochastic equation for the COM motion of a cluster, coupled to a stochastic process describing the time evolution of the instantaneous number of particles \(N(t)\) inside the cluster, thereby explicitly incorporating particle gain and loss.
We then validate and specialize the model by direct comparison with ABP simulations of isolated clusters held in stationary conditions (a single dense cluster coexisting with a  dilute phase). In these simulation, we show that \(N(t)\) is well described by a Gaussian process with mean \(N_0\), variance scaling as \(N_0^\beta\), and persistence time scaling as \(N_0^\kappa\), with $\beta=0.82$ and $\kappa=0.45$.
Solving the coupled equations analytically in the case of the Gaussian process for \(N(t)\), we find that the long-time diffusion coefficient \(D\) of the cluster is the sum of two contributions:
(i) the conventional term \(\propto N_0^{-1}\), arising from thermal noise and the cumulative effect of the active forces, and
(ii) an additional term originating from mass fluctuations, scaling as \(N_0^{-\delta}\) with \(\delta = 2 - \frac{2}{d} - \beta + \kappa\),
where \(d\) is the spatial dimension. 
Thus, anomalous scaling of \(D\) naturally emerges whenever the second term becomes dominant over the first one. The effective exponent depends on both the system dimensionality \(d\) and the scaling of the mass-process parameters  with the mean cluster size \(N_0\).
Direct comparison with the ABP simulations of Caporusso et al.~\cite{claudio_clusters_23} demonstrates good quantitative agreement: our model predicts \(D \sim N^{-\alpha}\) with \(\alpha = 0.63 \pm 0.06\).

\begin{figure}[t]
    \centering
    \includegraphics[width=0.65\textwidth]{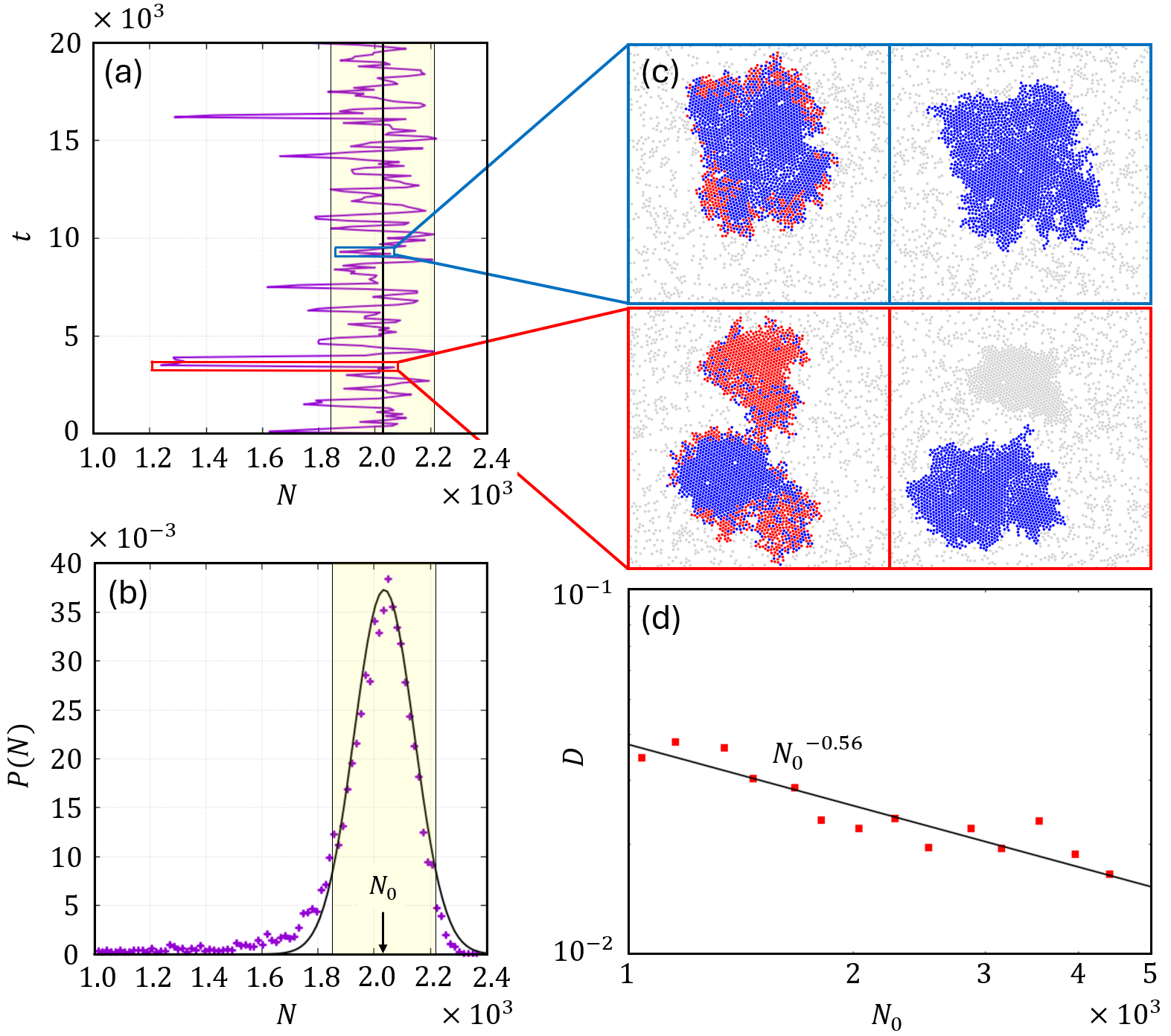}
    \caption{(a) Time evolution of a stationary cluster of $N(t)$ particles, taken from an ABPs simulation with a total of 3025 particles (cluster mean size \(N_0=2036\) as black line) and (b) corresponding probability distribution \(P(N)\), see SM for simulation details. The yellow shaded region marks the range where \(N\) fluctuates within $10\%$ of $N_0$; within this region \(P(N)\) is approximately Gaussian (black line fit in (b)). (c) Snapshots at two consecutive close times (left to right) with particles lost from  the left snapshots highlighted in red; top (bottom) row shows a mass-change event inside (outside) the yellow region of (a). 
    (d) Cluster's diffusion coefficient \(D\) versus mean cluster size \(N_0\) from ABP simulations (SM for details). The continuous black line indicates the fitted power law \(1.8\cdot N_0^{-0.56}\). }
    \label{fig:new_simsulations}
\end{figure}

\section{Phenomenology of ABP cluster motion from  numerical simulations}
In order to build a theoretical model describing the equation of motion of a single cluster of active particles, we first briefly illustrate the phenomenology of the cluster’s motion, reproducing results previously observed in molecular dynamics simulations of Active Brownian Particles (ABPs)~\cite{claudio_clusters_23}. Note that here we focus only on one specific case in which anomalous scaling was observed, i.e. fixed moderate active force and $N_0 \in [10^3, 5\cdot 10^3]$, while a complete characterization at different activity values is left for future work.
We report in the Supplementary Material (SM) a detailed description of the model, equations of motion, parameters used and simulation methodology. 

A collection of ABPs in certain conditions of density and activity undergoes MIPS\cite{Fily_2012,Render_2013,claudio_clusters_23,mips_abp_dumbells_2019, Cugliandolo_2017}, i.e. phase separates into a single cluster corresponding to the dense phase, and a dilute phase. 
Once formed, the cluster at stationarity exhibits fluctuations in its number of particles, 
$N(t)$, due to mass exchange with the dilute phase, while maintaining an average value $N_0$,
see Fig.~\ref{fig:new_simsulations}~(a) for a typical time evolution~\cite{claudio_clusters_23}. The mass distribution $P(N)$, shown in Fig.~\ref{fig:new_simsulations}~(b), indicates that typical fluctuations are Gaussian, resulting from particles randomly detaching from or attaching to the cluster boundary (top row of Fig.~\ref{fig:new_simsulations}~(c), showing two successive frames with particles detaching highlighted in red). However, rare events involving the attachment or detachment of large mass pieces produce a non-Gaussian tail (bottom row of Fig.~\ref{fig:new_simsulations}~(c)). Independently of the type of fluctuation event, any change in the number of particles induces a shift of the cluster's center of mass (COM), which in turn affects its motility (see also Fig.~SM1). 

In this work, we focus our analysis exclusively on the Gaussian fluctuations, which corresponds to neglecting fragmentation into  two or more clusters of comparable size. Such macroscopic fragmentation events, although present, are rare. When they occur, the identification of a main cluster and its center of mass becomes ambiguous. 
This choice also allows a direct comparison with previous results~\cite{claudio_clusters_23}, where only trajectory segments with $  N(t)  $ remaining within $10\%$ of $  N_0  $ (highlighted in yellow in Fig.~\ref{fig:new_simsulations}~(a) and (b)) were considered, corresponding to predominantly Gaussian fluctuations. Using the same procedure for the analysis of 
Analysing the ABPs simulation trajectories (see the details in the SM), we measured the mean-square displacement of the cluster's center of mass for varying $N_0$ (Fig.~SM2).
Fitting these curves, we obtain the diffusion coefficient $D$, Fig.~\ref{fig:new_simsulations}~(d).
The fit (continuos black line) with $D\sim N_0^{-\alpha}$ yields $\alpha=0.56\pm 0.07$, in good agreement with  $\alpha=0.5$ of Ref.~\cite{claudio_clusters_23}. Note that both the fluctuations described above and the anomalous scaling of the diffusion coefficient are directly caused by the particle activity. In contrast, systems of passive Brownian particles with attractive interactions do not exhibit these properties~\cite{claudio_clusters_23}.

\section{Theoretical model for the diffusion of cluster's center of mass}
Inspired by observations from ABP simulations, we now aim to construct a theoretical model capable of describing the evolution of a cluster's COM position due to mass exchange with its environment, starting from the dynamics of the individual particles composing the cluster.
For analytical simplicity, we consider active Ornstein–Uhlenbeck particles (AOUPs)~\cite{szamel2014self,caprini2018active,nguyen2021active,martin2021statistical,keta2024emerging,bonilla_2019}, where self-propulsion is modeled as Gaussian colored noise with finite persistence time. 
Note that, although activity in ABPs is represented by a constant-magnitude self-propulsion force whose orientation undergoes rotational diffusion, the parameters of the ABP and AOUP models can be directly mapped onto each other~\cite{caprini2022parental,sprenger_abp_aoup_2023}, \textcolor{black}{provided that the steady-state correlations of the self-propulsion velocity are matched. Although in the transient regime the self-propulsion behavior is different, AOUPs still display MIPS~\cite{cates_mips_2008}.}

Suppose we start from a cluster of $N$ AOUPs (formed through MIPS), with  each particle of diameter $\sigma$ and mass $m$ evolving in $d$-dimensions according to the following overdamped Langevin equation:

\begin{equation} \label{eq:abp_eq_continuous}
    \dot{\bm{r}}_{i}(t) = \bm{u}_{i}(t) +\sqrt{ 2D_{T}} \ \bm{\xi}_{i} (t) - \bm{\nabla}_{i} U \ \rm{,}
\end{equation} 
with $\bm{r}_{i}$,$\bm{u}_{i}$ the $i$-th particle's position and  self-propulsion velocity, $D_T=k_BT/m\gamma$ the translational diffusion coefficient, $\gamma$ and $T$ the friction coefficient per unit mass and the temperature of the thermal bath, respectively.  
$\bm{\xi}_i(t)$ is a $d$-dimensional Gaussian white noise with $\langle\xi_{i,a}(t)\rangle=0$ and $\langle\xi_{i,a}(t)\xi_{j,b}(t')\rangle=\delta_{ij}\delta_{ab}\delta(t-t')$, with $a,b=1,...,d$. $U$ represents a generic pairwise interaction potential, accounting for repulsive interactions between all particle pairs.
The self-propulsion  velocity $\bm{u}_{i}$ evolves as a coloured noise, according to the following equation:
\begin{equation} \label{eq:abp_eq_continuous1}
    \dot{\bm{u}}_{i}(t) = -\frac{\bm{u}_{i}(t)}{\tau_u}  + \sqrt{\frac{2D_u}{\tau_u^2}} \ \bm{\eta}_{i}(t) \ \rm{,}
\end{equation}
where $D_u$ quantifies the effective diffusion coefficient
of the particle, $\tau_u$ is the characteristic persistence time controlling the decay of $\bm{u}_{i}(t)$, $\bm{\eta}_i(t)$ is a white Gaussian noise with the same characteristics as $\bm{\xi}_i(t)$.  Note that the variance of the $\bm{u}_{i}$ process can be expressed in terms of the magnitude of the active force, $F_a$, using the following relation $D_u/\tau_u=(F_a/m\gamma)^2$~\cite{caprini2022parental}.
For initial stationary conditions of $\bm{u}_{i}$, mean and variance for this process read:
\begin{equation}
    \langle u_{a}(t)\rangle=0, \quad \langle u_{a}(t) u_{b}(t')\rangle = \delta_{ab} \frac{D_u}{\tau_u} e^{-|t'-t|/\tau_u} \ .
\end{equation}

We now consider that the number of particles inside the cluster, $N(t)$, changes over time as suggested by the simulations of  Fig.~\ref{fig:new_simsulations}~(a). 
Thus, at two consecutive times we have $N(t)$ and $N(t+\Delta t)=N(t)+\Delta N(t)$, where $\Delta N(t) > 0$ if particles have attached to the cluster and $\Delta N(t) < 0$ if they have detached.  Accordingly, the cluster's center of mass, defined as $\bm{r}_{CM}^{N(t)}(t) = \frac{1}{N(t)}\sum_{i=1}^{N(t)} \bm{r}_{i}(t)$ will displace as:  
\begin{equation}
    \label{eq:x_com_t+dt}
    \begin{aligned}
    \bm{r}_{CM}^{N(t+\Delta t)}(t+\Delta t) &= \frac{1}{N(t+\Delta t)} \sum_{i=1}^{N(t+\Delta t)} \bm{r}_{i}(t+\Delta t) \rm{.} 
    \end{aligned}
\end{equation}
We emphasize that in Eq.~\eqref{eq:x_com_t+dt} the sum runs over particle indices that may differ from those at time $t$ due to mass exchange with the environment.
We now assume $|\Delta N(t)|<N(t)$, and we impose $N(t)\ge 1$, as in simulations of Fig.~\ref{fig:new_simsulations}.
In this way, we can write a general relationship between $\bm{r}_{CM}^{N(t+\Delta t)}(t+\Delta t)$ and $\bm{r}_{CM}^{N(t)}(t)$, as

{\small
\begin{align} \label{eq:x_t+dt_total} 
    \bm{r}_{CM}^{N(t+\Delta t)}&(t+\Delta t)= \notag \\
    &= \frac{N(t)\bm{r}_{CM}^{N(t)}(t+\Delta t)}{N(t)+\Delta N(t)} + \frac{\Delta N(t)\bm{r}_{CM}^{\Delta N(t)}(t+\Delta t)}{N(t)+\Delta N(t)}.
\end{align}}
Here, $\bm{r}_{CM}^{\Delta N(t)}(t+\Delta t)$ is the center of mass of the attaching/detaching fragment, while $\bm{r}_{CM}^{
N(t)}(t+\Delta t)$ is the center of mass of the particles belonging to the cluster at time $t$, advanced by $\Delta t$ through a discretized version of Eqs.~\eqref{eq:abp_eq_continuous}-\eqref{eq:abp_eq_continuous1}:

{\small
\begin{equation}\label{eq:abp_eq_xn_discrete}
    \begin{cases}
        \bm{r}_{CM}^{N(t)}(t+\Delta t) = \bm{r}_{CM}^{N(t)}(t) + \bm{u}_{CM}^{N(t)}(t)\Delta t +\sqrt{ \frac{2D_T}{N(t)}} \Delta \bm{W}_{\xi}  \\
        \bm{u}_{CM}^{N(t)}(t+\Delta t) = \bm{u}_{CM}^{N(t)}(t) (1 - \frac{\Delta t}{\tau_u})+ \sqrt{\frac{2 D_u}{N(t)\tau_u^2}}\Delta \bm{W}_{\eta}, 
    \end{cases}   
\end{equation}}
with $\Delta \bm{W}_{\xi}$ and $\Delta \bm{W}_{\eta}$ random variables with Gaussian distribution $\mathcal{N}(0,\Delta t)$, and pair potential forces cancelling out.
Note that
{\small
\begin{align} \label{eq:corr_u_cm}
    \langle u_{CM,a}^{N(t)}(t) \rangle &= 0 \ , \notag \\
    \langle u_{CM,a}^{N(t)}(t) u_{CM,b}^{N(t')}(t') \rangle &= \frac{2D_u}{\tau_u^2} \int_0^t dt_1 \int_0^{t'} dt_2  e^{-\frac{t+t'-t_1-t_2}{\tau_u}} \cdot \notag \\
    &\quad \quad \cdot \langle \eta_a(t_1) \eta_b(t_2) \rangle \langle\frac{1}{\sqrt{N(t_1)N(t_2)}} \rangle \notag \\
    &= \delta_{ab}\langle \frac{1}{N}\rangle \frac{D_u}{\tau_u}  e^{-|t'-t|/\tau_u}, 
\end{align}}
where we used the fact that $N(t)$ and $\bm\eta$ are uncorrelated.
The average $\langle 1/N\rangle$ is intended over the cluster size distribution $P(N)$, assumed stationary.

\begin{figure}[t]
    \centering
    \includegraphics[width=0.65\textwidth]{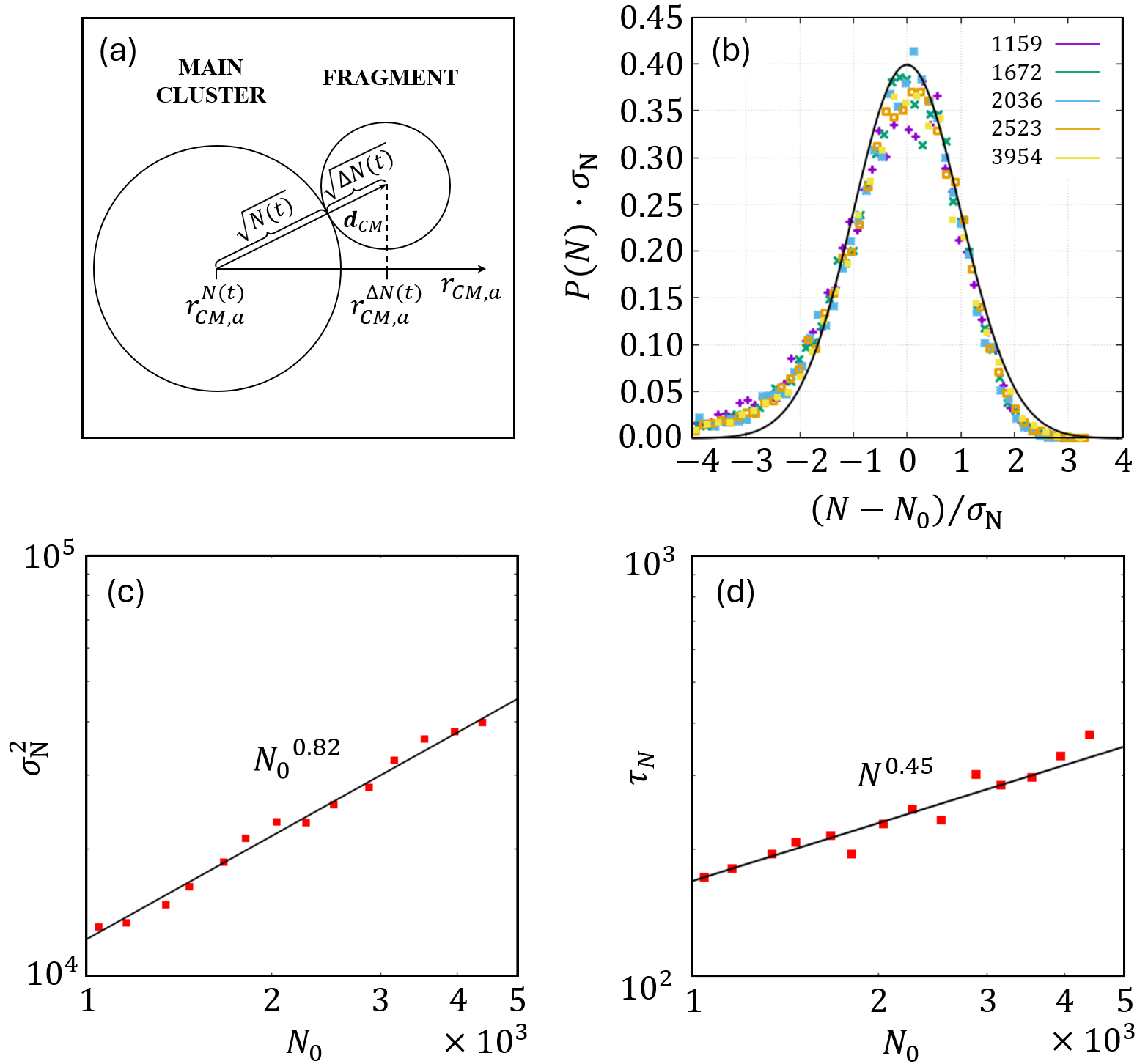}
    \caption{
    (a) Schematic of fragment attachment to the main cluster. The fragment (center-of-mass \(\mathbf{r}_{\rm CM}^{\Delta N}\)) and the main cluster (center-of-mass \(\mathbf{r}_{\rm CM}^{N}\)) are connected by the vector \(\mathbf{d}_{\rm cm}=|\mathbf{d}_{\rm cm}|\boldsymbol{\zeta}\), where \(\boldsymbol{\zeta}\) is a random unit vector drawn from the surface of a \(d\)-dimensional sphere and \(|\mathbf{d}_{\rm cm}|\) equals the sum of the two radii. Projections of the centers of mass along an arbitrary direction \(r_{\rm CM,a}\) are also shown.
    (b) Rescaled probability distributions \(P(N)\) from ABP simulations at different mean cluster sizes $N_0$. The solid line is the standard normal distribution \(\mathcal{N}(0,1)\).
    (c) Variance \(\sigma_N^2\) versus mean cluster size \(N_0\), extracted from the distributions in (b). The solid line is a power-law fit \(p N_0^\beta\) with \(\beta=0.82\pm0.03\) and \(p=40.4\pm 7.51\).
    (d) Persistence time \(\tau_N\) versus \(N_0\), obtained from the autocorrelation function of \(N(t)\) (see SM for details), together with the power-law fit \(cN_0^\kappa\) (solid line) with \(\kappa=0.45\pm0.05\) and \(c=7.20\pm3.76\).}
    \label{fig:new_model}
\end{figure}

We now assume that at time $t$ the group of $\Delta N(t)$ particles attaches to the main cluster of $N(t)$ particles or detaches from it along a random direction in the $d$-dimensional space, see the snapshots of Fig.~\ref{fig:new_simsulations}~(c) and scheme of Fig.~\ref{fig:new_model}~(a). This assumption is general, and can involve either Gaussian or atypical mass fluctuations observed for ABP. 
Thus, at time $t+\Delta t$, the center of mass of the two clusters are related through a vector $\bm{d}_{CM}$:
\begin{equation}
    \bm{r}_{CM}^{\Delta N(t)}(t+\Delta t)=\bm{r}_{CM}^{N(t)}(t+\Delta t)+|\bm{d}_{CM}|\bm\zeta,
\end{equation}
with $\bm\zeta$ is a random vector, uncorrelated in time, on a $d$-dimensional unit sphere \cite{hillier2009_spherical_vectors, Vershynin_2018}, such that the $a$-th component has distribution $P(\zeta_a)=\frac{\Gamma(d/2)}{\sqrt{\pi}\ \Gamma((d-1)/2)} (1-\zeta_a^2)^{(d-3)/2}$, with $\Gamma$ the gamma function, $\langle \zeta_a \rangle = 0$ and $\langle \zeta_a \zeta_b\rangle=\delta_{ab}/d$.

Moreover, since we assume the two clusters to be spherical, the length of $\bm{d}_{CM}$ can be set equal to the sum/subtraction of the radius of the two clusters for the attaching/detaching case, respectively (Fig.~\ref{fig:new_model}~(a)):
{\small
\begin{align*}
    |\bm{d}_{CM}| =\frac{\sigma}{2} \left( N(t)^{1/d} + sgn(\Delta N(t)) \cdot |\Delta N(t)|^{1/d}\right),
\end{align*}}
with $sng$ the sign function. 

Finally, the position of the cluster of $N(t+\Delta t)$ particles is obtained from Eqs.\eqref{eq:x_com_t+dt}-\eqref{eq:x_t+dt_total} as:
{\small 
\begin{align} \label{formula_finalee}
    &\bm{r}_{CM}^{N(t+\Delta t)}(t+\Delta t) = \bm{r}_{CM}^{N(t)}(t+\Delta t) + \notag \\
    &\quad \quad \quad +\frac{\Delta N(t) \big( N(t)^{\frac{1}{d}}  + sgn(\Delta N(t))|\Delta N(t)|^{\frac{1}{d}}\big)}{N(t)+\Delta N(t)} \frac{\sigma}{2} \bm\zeta .
\end{align}}
Eq.~\eqref{formula_finalee}, along with Eq.~\eqref{eq:abp_eq_xn_discrete}, can now be used to study the time evolution of the cluster's COM  with varying mass. 
Only the process of $N(t)$ remains to be specified. 

We write the equation of the velocity for the cluster's COM as $(\bm{r}_{CM}^{N(t+\Delta t)}(t+\Delta t)-\bm{r}_{CM}^{N(t)}(t))/\Delta t$, which in the limit $\Delta t \to 0$ becomes:
{\small
\begin{align}
    \bm{\dot r}_{CM}(t) = \bm{u}_{CM}^{N(t)}(t) +\sqrt{ \frac{2D_T}{N(t)}} \bm{\xi}(t)  + \chi (t) \bm\zeta, \label{eqrdot}
\end{align}}
with
{\small
\begin{equation} \label{eq:chi_gauss}
   \chi (t)= \frac{\sigma}{2} \frac{d N(t)}{dt}\frac{\big( N(t)^{1/d} + sgn(d N(t))|d N(t)|^{1/d}\big)}{N(t)+d N(t)} .
\end{equation}}
Note that $ \chi (t)$ and $\bm\zeta $ are independent processes. Moreover, depending on the underlying evolution of the process $N(t)$, $\chi$ can be simplified, in the case where $dN \ll N$, as:
{\small
\begin{equation}\label{chi}
   \chi (t)= \frac{\sigma}{2} \frac{dN}{dt} N(t)^{-1+\frac{1}{d}} .
\end{equation}}

The mean square displacement (MSD) of the $a$ component of $\bm{r}_{CM}$ in $d$ dimensions reads:
{\small
\begin{align}
    \text{MSD}(t) = \int_0^t dt_1 \int_0^t dt_2 \left<\Dot{r}_{CM,a}(t_1)\Dot{r}_{CM,a}(t_2)\right> \rm{.}
\end{align}}
Substituting Eq.~\eqref{eqrdot}, since the processes $u_{CM,a}$, $\xi_a$ and $\zeta_a$ are mutually independent, with zero average, and since $\xi_a$ and $\zeta_a$ are independent on the mass process $N$, we obtain:
{\small
\begin{align}
    \text{MSD}(t) &= \int_0^t dt_1 \int_0^t dt_2 \langle  u_{CM,a}^{N(t_1)}(t_1)u_{CM,a}^{N(t_2)}(t_2) \rangle \notag \\
    &\quad + \int_0^t dt_1 \int_0^t dt_2 2D_T \langle \frac{1}{\sqrt{N(t_1)N(t_2)}} \rangle \langle\xi_{a}(t_1) \xi_a(t_2) \rangle \notag \\
    &\quad + \int_0^t dt_1 \int_0^t dt_2  \langle \chi(t_1)\chi(t_2) \rangle\langle \zeta_a\zeta_a\rangle.
\end{align}}
Finally, substituting correlations and considering that $P(N)$ is stationary, we obtain:
{\small
\begin{align} \label{eq:msd_general}   
    \text{MSD}(t) &= 2 D_u \langle \frac{1}{N}\rangle \big[t-\tau_u(1-e^{-t/t_u})\big]+ 2D_T\langle\frac{1}{N}\rangle t \notag \\
    &\quad + \frac{1}{d}\int_0^t dt_1 \int_0^t dt_2  \langle \chi(t_1)\chi(t_2) \rangle.
\end{align}
}
Note how translational noise and active force contribute to the diffusion coefficient via a term proportional to $\langle 1/N\rangle$. 
In typical situations where the relative mass fluctuations are small, i.e. the mean cluster size $  N_0 = \langle N \rangle  $ greatly exceeds the standard deviation of its fluctuations, $  N_0 \gg \sigma_N  $, one can directly approximate $  \langle N^\lambda \rangle  \sim N_0^{\lambda}$. Consequently, we recover for the first two terms in Eq.~eq\ref{eq:msd_general}  the standard scaling $N_0^{-1}$ expected in the case of an active cluster with constant mass~\cite{claudio_clusters_23}. 
Conversely, the term proportional to $  \chi  $, stemming from center-of-mass shifts caused by particle attachment and detachment 
, gives rise to an additional contribution that can lead to anomalous scaling of the diffusion coefficient $  D  $.

\section{Diffusion coefficient for a mass Gaussian process}
In order to obtain an explicit expression for $D$, we must specify the underlying stochastic process governing the mass and evaluate averages with respect to the stationary mass distribution $P(N)$.
Here, we extract the properties of this distribution from a direct comparison with ABP simulations.
As shown in Fig.~\ref{fig:new_simsulations}~(b), in the region where $D$ is computed, $P(N)$ is well fitted by a Gaussian of mean $N_0$ and standard deviation $\sigma_N$.
The mean-shifted and variance-rescaled distributions are compared with the standard normal in Fig.~\ref{fig:new_model}~(b) for different $N_0$, while $\sigma_N^2$ as a function of $N_0$ is reported in 
Fig.~\ref{fig:new_model}~(c). We find that these values can be fitted by a power law, 
$\sigma_N^2 = p N_0^{\beta}$, with  $p=40.4\pm 7.51$ and $\beta = 0.82 \pm 0.03$, the latter smaller than the value $\beta = 1$ expected for typical mass fluctuations. To completely specify the Gaussian process, we also need to extract the 
 correlation time $\tau_N$  from exponential fits of the mass autocorrelation function (see SM); $\tau_N$ also follows  a power law $\tau_N = c N_0^{\kappa}$ with $c=7.20\pm3.76$, and $\kappa = 0.45 \pm 0.05$ (Fig.~\ref{fig:new_model}~(d)).

Now we can define the mass process as:
\begin{equation}\label{eq:g.1}
    \dot N(t) = - \frac{N(t)-N_0}{\tau_N} + \sqrt{\frac{2\sigma_N^2}{\tau_N}} \nu(t) \ ,
\end{equation}
where we assume the scaling $\sigma_N^2 = p N_0^{\beta}$ and $\tau_N = c N_0^{\kappa}$, and $\nu(t)$ is a white noise with the same properties as $\xi_a(t)$. 
The process has reflecting boundary conditions at $N=1$, in order to ensure $N\ge1$ at all times.

Finally, we observe $N_0 \gg \sigma_N$ across all cases, so that $\langle N^{\lambda}\rangle \sim N_0^{\lambda}$.

Using Eq.\eqref{eq:g.1}, we can now compute the last term in Eq.\eqref{eq:msd_general}:
{\small
\begin{align} \label{third_term_gaussian}
    \frac{1}{d}\int_0^t dt_1 \int_0^t &dt_2  \langle \chi(t_1)\chi(t_2) \rangle = \notag \\
    &= \frac{\sigma^2}{4d}\int_0^t dt_1 \int_0^t dt_2 \langle \frac{\dot N(t_1)\dot N(t_2)}{(N(t_1)N(t_2))^{1-\frac{1}{d}}} \rangle \notag\\
    &\simeq \frac{\sigma^2p}{d\ c} N_0^{\frac{2}{d}-2+\beta-\kappa} t \ \rm{,}
\end{align}}
where we used $\langle (N(t_1)N(t_2))^{1-\frac{1}{d}}\rangle \simeq N_0^{2-\frac{2}{d}}$ and $t\gg \tau_N$.  
Finally, approximating $\langle N^{-1}\rangle\simeq N_0^{-1}$ and $t\gg \tau_u$ we obtain the diffusion coefficient of the cluster's center of mass: 
\begin{equation}\label{coeffdifffinale}
    D=D_aN_0^{-1}+D_g N_0^{-\delta},
\end{equation}
with $D_a=D_u+D_T$ the diffusion coefficient of a single active particle,  $D_g=\frac{\sigma^2p}{2dc}$, and $\delta=2-2/d-\beta +\kappa$.
We first compare the theory with ABP simulations. In the latter, 
the single-particle diffusion coefficient  is $D_a = 8.33$ (SM), while the fits in Fig.~\ref{fig:new_model} directly yield $D_g = 1.4\pm 1.2$ and $\delta = 0.63 \pm 0.06$. 
In this regime, the scaling term proportional to $D_g$ dominates over the one proportional to $D_a$, leading to an overall scaling $D \sim N_0^{-\alpha}$ with $\alpha \approx \delta$. 
This prediction is in good quantitative agreement with the simulation results.

More in general, Eq.~\eqref{coeffdifffinale} reveals two contributions to $D$. The conventional term $\propto N_0^{-1}$ arises from the sum of active forces and thermal noise, as in standard Brownian systems. In addiction, we have an additional anomalous term $\propto N_0^{-\delta}$ which stems from  evaporation/condensation of particles at the cluster boundary. The exponent $\delta$ combines a geometric factor $2 - 2/d$ (stronger in low dimensions due to larger center-of-mass shifts upon mass change) with a dynamical factor $\kappa - \beta$, reflecting the anomalous scalings $\sigma_N^2 \sim N_0^\beta$ and $\tau_N \sim N_0^\kappa$ characteristic of the cluster's mass intrinsic process.
For example, in $d=1$ with standard mass fluctuations $\beta=1$ and no dependence on persistence time $\kappa=0$, we have $\delta=-1$: each attachment/detachment induces a large center-of-mass shift ($\sim N_0^{0.5}$), causing larger clusters to diffuse faster than smaller ones. For larger dimensions, this effect becomes progressively less relevant, since each attachment/detachment event affects only one out of the $  d  $ possible spatial directions.

The relative magnitude of $D_a$ and $D_g$ determines which of the two contributions in Eq.~\eqref{coeffdifffinale} dominates the scaling of the diffusion coefficient $D$. 
A crossover between the two regimes occurs at a characteristic cluster size $N_c$ defined by
$D_a N_c^{-1} \approx D_g N_c^{-\delta}$, 
or equivalently
\begin{equation}
    N_c \approx \left( \frac{D_a}{D_g} \right)^{\frac{1}{1-\delta}}.
\end{equation}

The behavior depends on whether $\delta < 1$ or $\delta > 1$:
if $\delta < 1$, then for small clusters ($N_0 \ll N_c$) the classical term dominates and $D \sim N_0^{-1}$, while for large clusters ($N_0 \gg N_c$) the anomalous evaporation/condensation term takes over and $D \sim N_0^{-\delta}$.
If $\delta > 1$, the opposite occurs: $D \sim N_0^{-\delta}$ at small $N_0 \ll N_c$, and $D \sim N_0^{-1}$ at large $N_0 \gg N_c$. When $\delta = 1$, $D \sim N_0^{-1}$ holds across all cluster sizes.

\begin{figure}[t]
    \centering
    \includegraphics[width=0.75\textwidth]{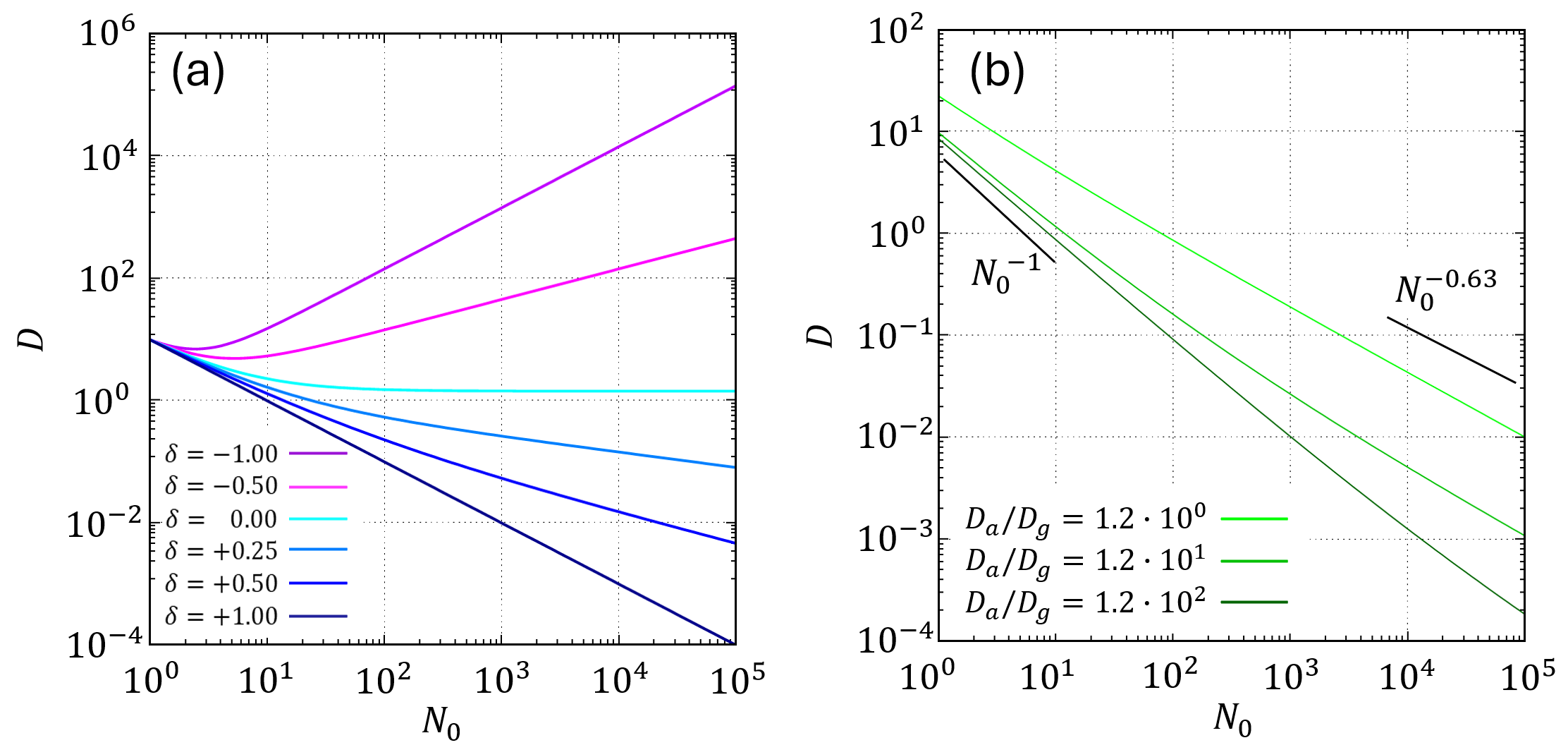}
    \caption{Diffusion coefficient \(D\) versus mean cluster size \(N_0\), as given by Eq.~\eqref{coeffdifffinale}, using \(D_a=8.33\), \(d=2\) (same value as ABPs simulations, see SM). (a) Fixed \(D_g=1.4\) (from fitted ABPs simulation values); and different values of \(\delta\). (b) Fixed \(\delta=0.63\) and different values of \(D_g\).}

    \label{fig:new_theoretical}
\end{figure}

As an example, in Fig.~\ref{fig:new_theoretical}~(a) we show $D(N_0)$ for $d=2$ with ABP-like parameters (in the caption) and varying $\delta \in [-1,1]$. Except when $\delta=1$ ($D \sim N_0^{-1}$ everywhere), a crossover from $D \sim N_0^{-1}$ to $D \sim N_0^{-\delta}$ occurs at $N_c$, which decreases with increasing $\delta$.
Fixing $\delta=0.63$ and $D_a$, and increasing $D_g$  shifts the crossover to lower $N_0$ (Fig.~\ref{fig:new_theoretical}~(b)).

We also consider, in the SM, a two-state mass process allowing only two discrete values for the cluster mass --- a scenario closer to large-scale fragmentation or attachment events (Fig.~\ref{fig:new_simsulations}~(c)). 
Even in this case, the diffusion coefficient follows precisely the same scaling as in Eq.~\ref{coeffdifffinale}. 
This confirms the robustness of the result: the anomalous scaling of $D$ is governed mainly by the exponents $\beta$ and $\kappa$ (describing mass variance and correlation time vs.\ mean mass) and by system dimensionality $d$, rather than by fine details of the mass fluctuation mechanism.

\section{Discussion} \label{conclusion}

In this work, we develop a theoretical model to describe the time evolution and diffusion properties of the center of mass (COM) of a cluster of active particles. 
The model is motivated by numerical observations of active Brownian particle (ABP) clusters in the stationary regime following complete motility-induced phase separation (MIPS). 
These clusters maintain a roughly constant average mass $N_0$ while continuously exchanging particles with the surrounding gas. Their center of mass diffuses with an anomalous scaling of the diffusion coefficient $D \sim N_0^{-\alpha}$, with fitted $\alpha=0.56\pm 0.07$, compatible with the value $0.5$ found in Ref.~\cite{claudio_clusters_23}.

Starting from the individual particle dynamics, we derive a general equation of motion for the cluster COM in $d$ dimensions, explicitly accounting for the stochastic mass process $N(t)$. 
For a Gaussian  process $\mathcal{N}(N_0, \sigma_N^2)$---consistent with ABP simulation results---we analytically compute the long-time diffusion coefficient. 
Simulations reveal anomalous scaling in the mass statistics: the variance scales as $\sigma_N^2 = p N_0^{\beta}$ and the persistence (correlation) time as $\tau_N = c N_0^{\kappa}$, with fitted exponents $\beta = 0.82$ and $\kappa = 0.45$.
This leads to the general expression
$D = D_a N_0^{-1} + D_g N_0^{-\delta}$,
where $D_a$ is the single-particle active diffusion coefficient, $D_g = \sigma^2 p / (2 d c)$, and 
$\delta = 2 - 2/d - \beta + \kappa$.

Thus, the diffusion coefficient contains two distinct contributions: (i) a conventional term $\propto N_0^{-1}$ arising from the summation of particles stochastic noises and active forces and (ii) an anomalous term $\propto N_0^{-\delta}$ that originates from mass-induced shifts of the center of mass. \textcolor{black}{For the conventional term, note that a similar $N_0^{-1}$ dependence is also expected in the case of ABPs~\cite{claudio_clusters_23}, even though the self-propulsion process is different from the one of AOUPs.} The anomalous one combines a geometric factor ($2 - 2/d$), which becomes more important in lower dimensions) with a dynamical factor ($\kappa - \beta$) reflecting the anomalous  dependence of mass fluctuations and their temporal correlations from the cluster's size. 
The ratio $D_a/D_g$ controls the emergence of an anomalous regime over the typical one.

Several directions remain open for future investigation. 
First, it would be valuable to use the same theoretical framework to characterize why in ABPs there is a crossover between $D \sim N_0^{-0.5}$ and $D \sim N_0^{-1}$ for increasing $N_0$~\cite{claudio_clusters_23}, as well as other values and types of activity.
Second, the present model relies on a Gaussian approximation for the mass process, which captures small evaporation–condensation events well but neglects rarer, larger mass jumps. 
Future extensions should incorporate the non-Gaussian tail of the mass distribution observed in ABP simulations, particularly by accounting for infrequent but significant attachment/detachment events of large fragments. 
Third, it would be interesting to explore the regime where mass fluctuations become comparable to the mean ($\sigma_N \sim N_0$).  
Finally, the current model assumes a stationary cluster with fixed average mass $N_0$. 
An important generalization would be to non-stationary situations, such as the early or intermediate stages of motility-induced phase separation where the dense phase is still growing~\cite{caporusso_micromacro_2020,claudio_clusters_23}.

\section{Acknowledgment}
We thank support by MIUR via the projects 
PRIN 2020/PFCXPE, 
PRIN 2022/HNW5YL, 
PRIN PNRR/P20222B5P9 and 
Quantum Sensing and Modeling for One-Health (QuaSiModO).

\bibliographystyle{unsrt}
\bibliography{references}

@book{dhont1996introduction,
  author    = {Jan K. G. Dhont},
  title     = {An Introduction to Dynamics of Colloids},
  volume    = {2},
  publisher = {Elsevier},
  year      = {1996}
}

@article{auge2009nmr,
  author  = {S. Augé and P. O. Schmit and C. A. Crutchfield and others},
  title   = {NMR measure of translational diffusion and fractal dimension. Application to molecular mass measurement},
  journal = {ACS Publications},
  volume  = {113},
  pages   = {1914--1918},
  year    = {2009}
}

@book{book_theory_pol_dynamics,
  author    = {Masao Doi and Samuel Frederick Edwards},
  title     = {The Theory of Polymer Dynamics},
  volume    = {73},
  publisher = {Oxford University Press},
  year      = {1988}
}

@article{Winkler_active_review,
  author  = {R. G. Winkler and J. Elgeti and G. Gompper},
  title   = {Active Polymers — Emergent Conformational and Dynamical Properties: A Brief Review},
  journal = {Journal of the Physical Society of Japan},
  volume  = {86},
  pages   = {101014},
  year    = {2017}
}

@article{Bianco_active_polymers,
  author  = {V. Bianco and E. Locatelli and P. Malgaretti},
  title   = {Globally and locally active polymers: shape, structure, and dynamics},
  journal = {Physical Review Letters},
  volume  = {121},
  pages   = {217802},
  year    = {2018}
}

@article{bonn_phase_separation_prl,
  author  = {A. Deblais and A. C. Maggs and D. Bonn and S. Woutersen},
  title   = {Phase Separation by Entanglement of Active Polymerlike Worms},
  journal = {Phys. Rev. Lett.},
  volume  = {124},
  number  = {20},
  pages   = {208006},
  year    = {2020}
}

@article{Marchetti2013,
  author  = {M. C. Marchetti and J. F. Joanny and S. Ramaswamy and T. B. Liverpool and J. Prost and M. Rao and R. Aditi Simha},
  title   = {Hydrodynamics of soft active matter},
  journal = {Rev. Mod. Phys.},
  volume  = {85},
  pages   = {1143},
  year    = {2013}
}

@article{Bechinger2016,
  author  = {C. Bechinger and R. Di Leonardo and H. Löwen and C. Reichhardt and G. Volpe and G. Volpe},
  title   = {Active particles in complex and crowded environments},
  journal = {Rev. Mod. Phys.},
  volume  = {88},
  pages   = {045006},
  year    = {2016}
}

@article{Eldeti_2015,
  author  = {J. Elgeti and R. G. Winkler and G. Gompper},
  title   = {Physics of microswimmers—Single particle motion and collective behavior: A review},
  journal = {Rep. Prog. Phys.},
  volume  = {78},
  pages   = {056601},
  year    = {2015}
}

@article{Cates_2015,
  author  = {M. Cates and J. Tailleur},
  title   = {Motility-induced phase separation},
  journal = {Annu. Rev. Condens. Matter Phys.},
  volume  = {6},
  pages   = {219},
  year    = {2015}
}

@article{Fily_2012,
  author  = {Y. Fily and M. C. Marchetti},
  title   = {A thermal phase separation of self-propelled particles with no alignment},
  journal = {Phys. Rev. Lett.},
  volume  = {108},
  pages   = {235702},
  year    = {2012}
}

@article{Gonnella_2015,
  author  = {G. Gonnella and D. Marenduzzo and A. Suma and A. Tiribocchi},
  title   = {Motility-induced phase separation and coarsening in active matter},
  journal = {C. R. Phys.},
  volume  = {16},
  pages   = {316},
  year    = {2015}
}

@article{Render_2013,
  author  = {G. Redner and M. Hagan and A. Baskaran},
  title   = {Structure and dynamics of a phase-separating active colloidal fluid},
  journal = {Phys. Rev. Lett.},
  volume  = {110},
  pages   = {055701},
  year    = {2013}
}

@article{Cugliandolo_2017,
  author  = {L. Cugliandolo and P. Digregorio and G. Gonnella and A. Suma},
  title   = {Phase coexistence in two-dimensional passive and active dumbbell systems},
  journal = {Phys. Rev. Lett.},
  volume  = {119},
  pages   = {268002},
  year    = {2017}
}

@article{meakin_clusters_93,
  author  = {P. Meakin and A. T. Skjeltorp},
  title   = {Application of experimental and numerical models to the physics of multiparticle systems},
  journal = {Advances in Physics},
  volume  = {42},
  number  = {1},
  pages   = {1--127},
  year    = {1993}
}

@article{beatrici_clusters_PRE17,
  author  = {C. P. Beatrici and R. M. C. de Almeida and L. G. Brunnet},
  title   = {Mean-cluster approach indicates cell sorting time scales are determined by collective dynamics},
  journal = {Phys. Rev. E},
  volume  = {95},
  number  = {3},
  pages   = {032402},
  year    = {2017}
}

@article{claudio_clusters_23,
  author  = {C. B. Caporusso and L. F. Cugliandolo and P. Digregorio and G. Gonnella and D. Levis and A. Suma},
  title   = {Dynamics of Motility-Induced Clusters: Coarsening beyond Ostwald Ripening},
  journal = {Phys. Rev. Lett.},
  volume  = {131},
  number  = {6},
  pages   = {068201},
  year    = {2023}
}

@article{lutz2024anomalous,
  author  = {Aline Lütz and Carine Beatrici and Lívio Amaral and Leonardo Brunnet},
  title   = {Anomalous mass dependency in Hydra endoderm cell cluster diffusion},
  year    = {2024}
}

@article{Cottin-Bizonne_2018,
  author  = {F. Ginot and I. Theurkauff and F. Detcheverry and C. Ybert and C. Cottin-Bizonne},
  title   = {Aggregation-fragmentation and individual dynamics of active clusters},
  journal = {Nature Communications},
  volume  = {9},
  number  = {1},
  year    = {2018}
}

@article{hanggi_clusters_PRE95,
  author  = {J. Łuczka and P. Hänggi and A. Gadomski},
  title   = {Diffusion of clusters with randomly growing masses},
  volume  = {51},
  number  = {6},
  ournal = {Phys. Rev. E},
  pages   = {5762--5769},
  year    = {1995}
}

@article{rajesh2002aggregate,
  author  = {R. Rajesh and D. Das and B. Chakraborty and M. Barma},
  title   = {Aggregate formation in a system of coagulating and fragmenting particles with mass-dependent diffusion rates},
  journal = {Phys. Rev. E},
  volume  = {66},
  pages   = {056104},
  year    = {2002}
}

@article{seno_superstatistics_PRX17,
  author  = {A. V. Chechkin and F. Seno and R. Metzler and I. M. Sokolov},
  title   = {Brownian yet Non-Gaussian Diffusion: From Superstatistics to Subordination of Diffusing Diffusivities},
  journal = {Phys. Rev. X},
  volume  = {7},
  number  = {2},
  pages   = {021002},
  year    = {2017}
}

@article{hitchhiker_model,
  author  = {M. Hidalgo-Soria and E. Barkai},
  title   = {Hitchhiker model for Laplace diffusion processes},
  journal = {Phys. Rev. E},
  volume  = {102},
  pages   = {012109},
  year    = {2020}
}

@article{baldovin_polymerization_frontiers19,
  author  = {F. Baldovin and E. Orlandini and F. Seno},
  title   = {Polymerization Induces Non-Gaussian Diffusion},
  journal = {Frontiers in Physics},
  year    = {2019}
}

@article{nguyen2021active,
  author  = {G. H. P. Nguyen and R. Wittmann and H. Löwen},
  title   = {Active Ornstein-Uhlenbeck model for self-propelled particles with inertia},
  journal = {Journal of Physics: Condensed Matter},
  volume  = {34},
  number  = {3},
  pages   = {035101},
  year    = {2021}
}

@article{szamel2014self,
  title={Self-propelled particle in an external potential: Existence of an effective temperature},
  author={Szamel, Grzegorz},
  journal={Physical Review E},
  volume={90},
  number={1},
  pages={012111},
  year={2014},
  publisher={APS}
}

@article{caprini2018active,
  title={Active particles under confinement and effective force generation among surfaces},
  author={Caprini, Lorenzo and Marconi, Umberto Marini Bettolo},
  journal={Soft matter},
  volume={14},
  number={44},
  pages={9044--9054},
  year={2018},
  publisher={Royal Society of Chemistry}
}

@article{keta2024emerging,
  title={Emerging mesoscale flows and chaotic advection in dense active matter},
  author={Keta, Yann-Edwin and Klamser, Juliane U and Jack, Robert L and Berthier, Ludovic},
  journal={Physical Review Letters},
  volume={132},
  number={21},
  pages={218301},
  year={2024},
  publisher={APS}
}

@article{martin2021statistical,
  author  = {D. Martin and J. O'Byrne and M. E. Cates and others},
  title   = {Statistical mechanics of active Ornstein-Uhlenbeck particles},
  journal = {Physical Review E},
  volume  = {103},
  number  = {3},
  pages   = {032607},
  year    = {2021}
}

@article{caprini2022parental,
  title={The parental active model: A unifying stochastic description of self-propulsion},
  author={Caprini, Lorenzo and Sprenger, Alexander R and L{\"o}wen, Hartmut and Wittmann, Ren{\'e}},
  journal={The Journal of Chemical Physics},
  volume={156},
  number={7},
  year={2022},
  publisher={AIP Publishing}
}

@article{cates_mips_2008,
  author  = {J. Tailleur and M. E. Cates},
  title   = {Statistical Mechanics of Interacting Run-and-Tumble Bacteria},
  journal = {Phys. Rev. Lett.},
  volume  = {100},
  number  = {21},
  pages   = {218103},
  year    = {2008}
}

@article{bialke_lowen_mips_2013,
  author  = {Julian Bialké and Hartmut Löwen and Thomas Speck},
  title   = {Microscopic theory for the phase separation of self-propelled repulsive disks},
  journal = {Europhys. Lett.},
  volume  = {103},
  number  = {3},
  pages   = {30008},
  year    = {2013}
}

@article{digregorio_mips_2018,
  author  = {Pasquale Digregorio and Demian Levis and Antonio Suma and Leticia F. Cugliandolo and Giuseppe Gonnella and Ignacio Pagonabarraga},
  title   = {Full Phase Diagram of Active Brownian Disks: From Melting to Motility-Induced Phase Separation},
  journal = {Phys. Rev. Lett.},
  volume  = {121},
  number  = {9},
  pages   = {098003},
  year    = {2018}
}

@article{mips_abp_dumbells_2019,
  author  = {P. Digregorio and D. Levis and A. Suma and L. F. Cugliandolo and G. Gonnella and I. Pagonabarraga},
  title   = {2D melting and motility induced phase separation in Active Brownian Hard Disks and Dumbbells},
  journal = {Journal of Physics: Conference Series},
  volume  = {1163},
  number  = {1},
  pages   = {012073},
  year    = {2019}
}

@article{hillier2009_spherical_vectors,
  author  = {Grant Hillier and Raymond Kan and Xiaolu Wang},
  title   = {Computationally efficient recursions for top-order invariant polynomials with applications},
  journal = {Cambridge University Press},
  volume  = {1163},
  number  = {1},
  pages   = {211--242},
  year    = {2009}
}

@book{Vershynin_2018,
  author    = {R. Vershynin},
  title     = {High-Dimensional Probability: An Introduction with Applications in Data Science},
  publisher = {Cambridge University Press},
  year      = {2018}
}

@article{caporusso_micromacro_2020,
  title = {Motility-Induced Microphase and Macrophase Separation in a Two-Dimensional Active Brownian Particle System},
  author = {Caporusso, Claudio B. and Digregorio, Pasquale and Levis, Demian and Cugliandolo, Leticia F. and Gonnella, Giuseppe},
  journal = {Phys. Rev. Lett.},
  volume = {125},
  issue = {17},
  pages = {178004},
  numpages = {6},
  year = {2020},
  month = {Oct},
  publisher = {American Physical Society},
  doi = {10.1103/PhysRevLett.125.178004},
}

@article{bonilla_2019,
  title={Active ornstein-uhlenbeck particles},
  author={Bonilla, Luis L},
  journal={Physical Review E},
  volume={100},
  number={2},
  pages={022601},
  year={2019},
  publisher={APS}
}

@article{sprenger_abp_aoup_2023,
  title={Dynamics of active particles with translational and rotational inertia},
  author={Sprenger, Alexander R and Caprini, Lorenzo and L{\"o}wen, Hartmut and Wittmann, Ren{\'e}},
  journal={Journal of Physics: Condensed Matter},
  volume={35},
  number={30},
  pages={305101},
  year={2023},
  publisher={IOP Publishing}
}

\end{document}